\documentclass[conference]{IEEEtran}
\IEEEoverridecommandlockouts

\usepackage[draft]{pgf}
\usepackage{cite}
\usepackage{algorithm} 
\usepackage{algpseudocode} 
\usepackage{graphics} 
\usepackage{color}
\usepackage{amsmath, bm} 
\usepackage{amssymb}  
\usepackage{upgreek}
\usepackage{caption}
\usepackage{subcaption}
\usepackage{lipsum}
\def\BibTeX{{\rm B\kern-.05em{\sc i\kern-.025em b}\kern-.08em
    T\kern-.1667em\lower.7ex\hbox{E}\kern-.125emX}}
\begin{document}

\title{\LARGE \bf
Efficient Modeling of Morphing Wing Flight\\ Using Neural Networks and Cubature Rules
}

\author{Paul Ghanem, Yunus Bicer, Deniz Erdogmus, Alireza Ramezani
\thanks{$^{1}$The authors are with the Department of Electrical and Computer Engineering, Northeastern University, Boston, MA, USA. Emails: \{ghanem.p,
bicer.y a.ramezani,\}@northeastern.edu;\{erdogmus@ece.neu.edu\}}%
}

\maketitle

\begin{abstract}

Fluidic locomotion of flapping Micro Aerial Vehicles (MAVs) can be very complex, particularly when the rules from insect flight dynamics (fast flapping dynamics and light wings) are not applicable. In these situations, widely used averaging techniques can fail quickly. The primary motivation is to find efficient models for complex forms of aerial locomotion where wings constitute a large part of body mass (i.e., dominant inertial effects) and deform in multiple directions (i.e., morphing wing).

In these systems, high degrees of freedom yields complex inertial, Coriolis, and gravity terms. We use Algorithmic Differentiation (AD) and Bayesian filters computed with cubature rules conjointly to quickly estimate complex fluid-structure interactions. In general, Bayesian filters involve finding complex numerical integration (e.g., find posterior integrals). Using cubature rules to compute Gaussian-weighted integrals and AD, we show that the complex multi-degrees-of-freedom dynamics of morphing MAVs can be computed very efficiently and accurately. Therefore, our work facilitates closed-loop feedback control of these morphing MAVs.

\end{abstract}


\section{Introduction}

The dynamic modeling of mechanical and bio-inspired robotic systems is often conducted using ordinary differential equations (ODE) that describe the evolution of states such as positions and orientations over time. These models are obtained based on known physics of the mechanical systems and generally many interconnected processes inside of the described system are not known and cannot be modeled using ODE. For this reason, machine learning methods that identify the whole ODE\cite{ogunmolu_nonlinear_2016,lu_robust_1998} and physics-informed machine learning methods that fill the gaps in models based on ODE framework have emerged\cite{raissi_physics-informed_2019,stiasny_physics-informed_2021,qian_physics_2019}

Nevertheless, these machine learning methods are limited to offline training since they are trained using backpropagation and gradient descent methods. As a result, their training methods require huge data sets to converge for a solution which limits their time performance. The time of training particularly becomes important if these models are meant to be used in conjunction with fast inner-loop closed-loop feedback. This particularly becomes very problematic in robotic platforms that have highly nonlinear and continuous interactions with their fluidic environment, such as aerial and aquatic locomotion. While the application of machine learning is expanding in these disciplines, yet it is too early to make conclusive decision about their suitability for real-time closed-loop feedback because their accuracy and computation overhead is unknown.  

In other applications related to robot locomotion, often neural networks are trained around a local attractor with no possibility of updating neural networks weights in an online fashion. To enable online training, Bayesian approaches for training neural networks have emerged \cite{noauthor_training_nodate,wang_convergence_2011,safarinejadian_predict_2013}. However, the application of these Bayesian approaches remained limited.

\begin{figure}[!t]
    \centering
    \includegraphics[width=0.9\linewidth]{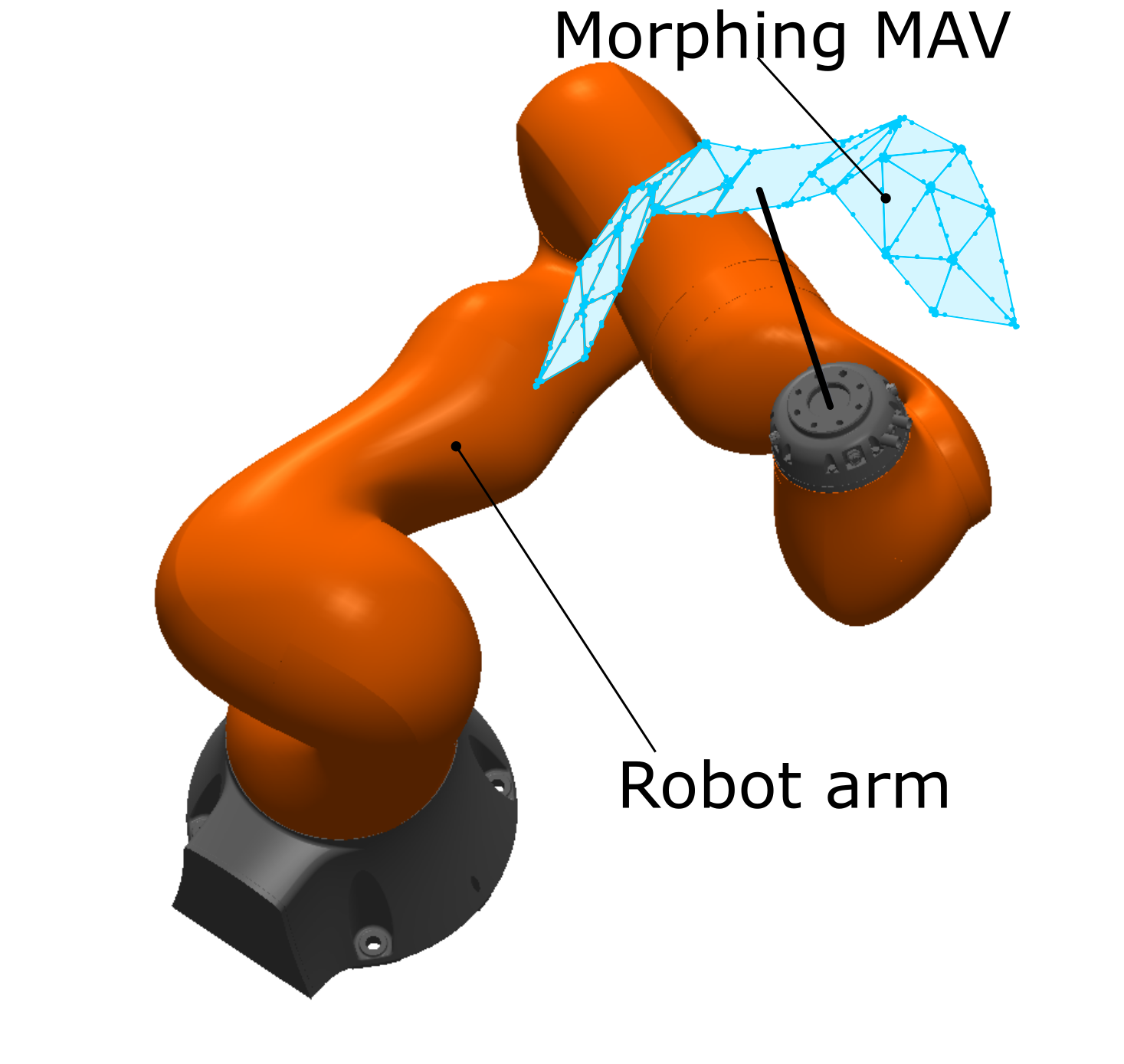}
    \caption{Shows a morphing MAV attached to a robot arm as a way of systematically studying fluid-structure interaction in the MAV.}
    \label{fig:cover_img}
\end{figure}

In this paper, we attempt to capitalize on an ongoing research project at Northeastern University, called \textit{Aerobat} \cite{sihite_computational_2020}, to explore novel methods of estimating complex robotic models using machine learning and neural net policies. Briefly speaking, Aerobat is a morphing Micro Aerial Vehicle (MAV) and its flight dynamics contains hard-to-model components involving complex fluid-structure interactions. 

The objective of this paper is to boost our efforts in understanding Aerobat's complex dynamics with a platform shown in Fig.~\ref{fig:cover_img}. Using the platform, it is possible to define any desired trajectories for the robot manipulator and Aerobat which the former defines flight path and the latter defines robot inputs. Then, the goal is to use the robot's Lagrangian model and measurements from onboard sensors to estimate models that realistically capture fluid-structure interactions in Aerobat. 

To do this, we use Algorithmic Differentiation (AD) and, inspired by \cite{arasaratnam_cubature_2009-1}, use Kalman filters that employ cubature approximation rules to numerically compute Gaussian-based integrals. As a result, we show that we can efficiently train the physics-informed, feed-forward neural network that captures Aerobat's dynamics.

This work is organized as follows. After briefly introducing NU's Aerobat, we drive the equations of motion using the method of Lagrange. Next, we describe two numerical approaches based on AD and cubature rules to extract efficient numerical models suitable for model-based nonlinear control of Aerobat. Our algorithmic computations are developed based on CasADi toolbox which is briefly described. The bulk of this works is focused on aerodynamic force estimation in our platform. We conclude this work with final remarks and simulation results.   

\section{Brief Overview of Northeastern Morphing MAV}

\begin{figure}[!t]
    \centering
    \includegraphics[width=0.9\linewidth]{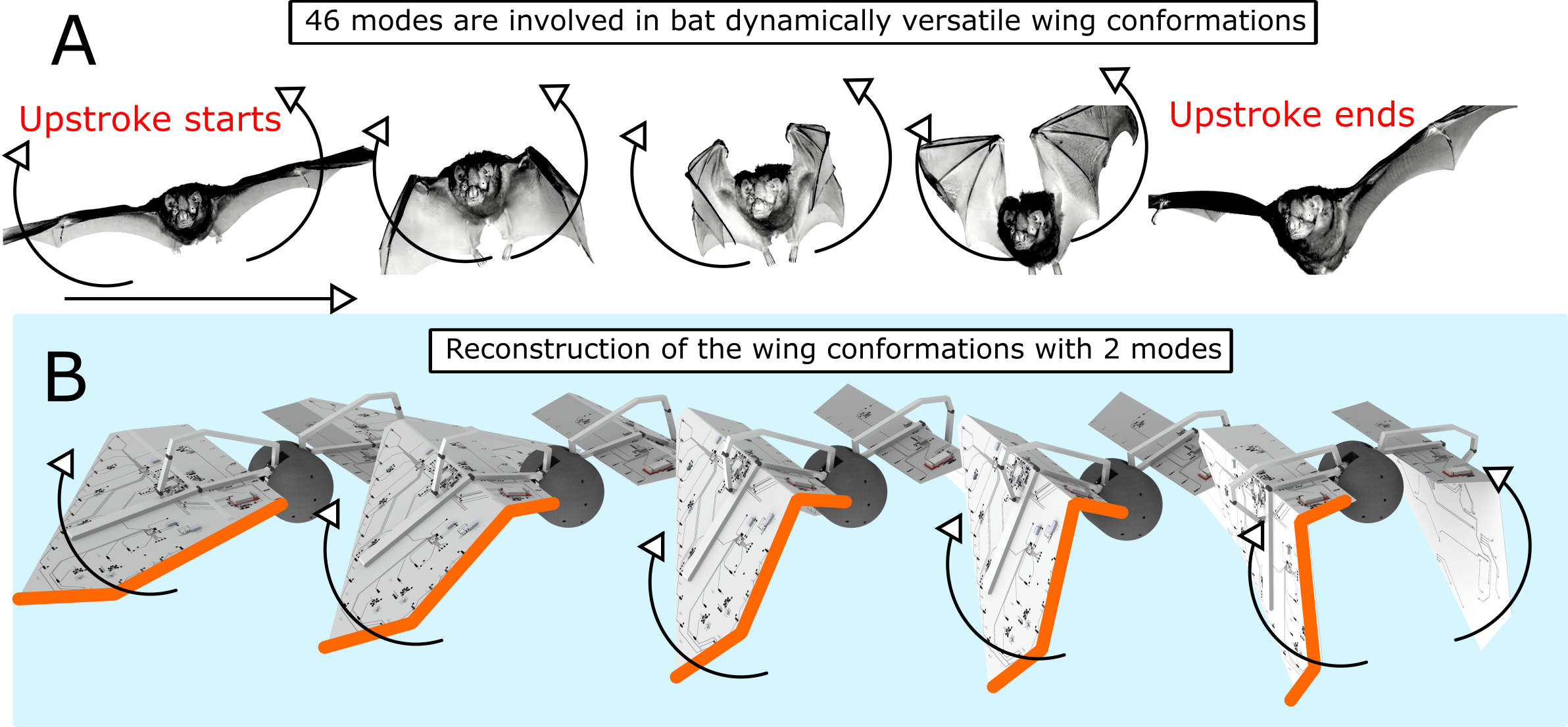}
    \caption{Shows Northeastern University's Aerobat \cite{sihite_computational_2020}, which is a morphing MAV.}
    \label{fig:aerobat}
\end{figure}

The Northeastern University's \textit{Aerobat}, which is a morphing MAV shown in Fig. \ref{fig:aerobat}, weighs less than 20 gr with a maximum wingspan of 32 cm and is capable of concurrently mobilizing 14 active linkage DOF in gait cycles shorter than 200 milliseconds. We have developed this platform (and the developement is still ongoing) to study control and actuation frameworks that can help us design future morphing MAVs.

\section{Morphing MAV Model}

Here, we describe the equations of motion which will be used to obtain the simulation results reported in this paper. The considered flight dynamics assume interconnected inertial dynamics and aerodynamic subsystem which is expected from these systems as they do not follow insect flight dynamics. The dynamics, as will be shown below, is affine-in-control which is less common to see in flapping wing flight. 

Let $q_u = [q_{x},q_{y},q_{z},p_{bx},p_{by},p_{bz}]^\top$ be the underactuated (passive) coordinates (body position and orientation parameterized with Euler angles) and $q_a$ be the actuated (active) coordinate vectors (wing joint angles). These underactuated and actuated coordinates result in a system defined by the configuration variable vector $q = [q_u^\top, q_a^\top]^\top$ and the state vector $x = [q^\top, \dot q^\top]^\top$.

Next, the equations of motion are derived using the Euler-Lagrange formalism after obtaining the system's Lagrangian $\mathcal{L}\in\mathbb{R}$. In obtaining the Lagrangian functional, the kinetic (translational and rotational) and potential energy led by the distributed mass from each wing segment is considered which for saving space explaining them is overlooked in this paper. Then the equations of motion are derived using the general formulation 

\begin{equation}
\frac{d}{dt} \frac{\partial \mathcal{L}}{\partial \dot q}  - \frac{\partial \mathcal{L}}{\partial q} = u_{gen}
\label{eq:general-EL-eq}    
\end{equation}

\noindent where $u_{gen}$ is the sum of the generalized forces. This formulation can be expanded into the following form:

\begin{equation}
\begin{aligned}
    \quad D(q) \ddot q + C(q, \dot q)\dot q + G(q) &= B_1(q) \, u_1 + B_2(q) \, u_2,
\end{aligned}
\label{eq:lagrangian_formulation}
\end{equation}

\noindent where $D(q)$ denotes the mass inertia matrix, $C(.)$ contains the Coriolis terms, $G(.)$ is the gravity vector, $u_1$ are the joint torques, $B_1(q) = \left(\partial q_a/\partial q\right)^\top$ maps joint torques and $B_2$ (similar to $B_1$) maps external forces to the generalized coordinates in the system. In Eq.~\ref{eq:lagrangian_formulation}, the aerodynamic contributions $u_2$ are considered as the second input to the system. 

\begin{figure*}[!t]
    \centering
    \includegraphics[width=0.9\linewidth]{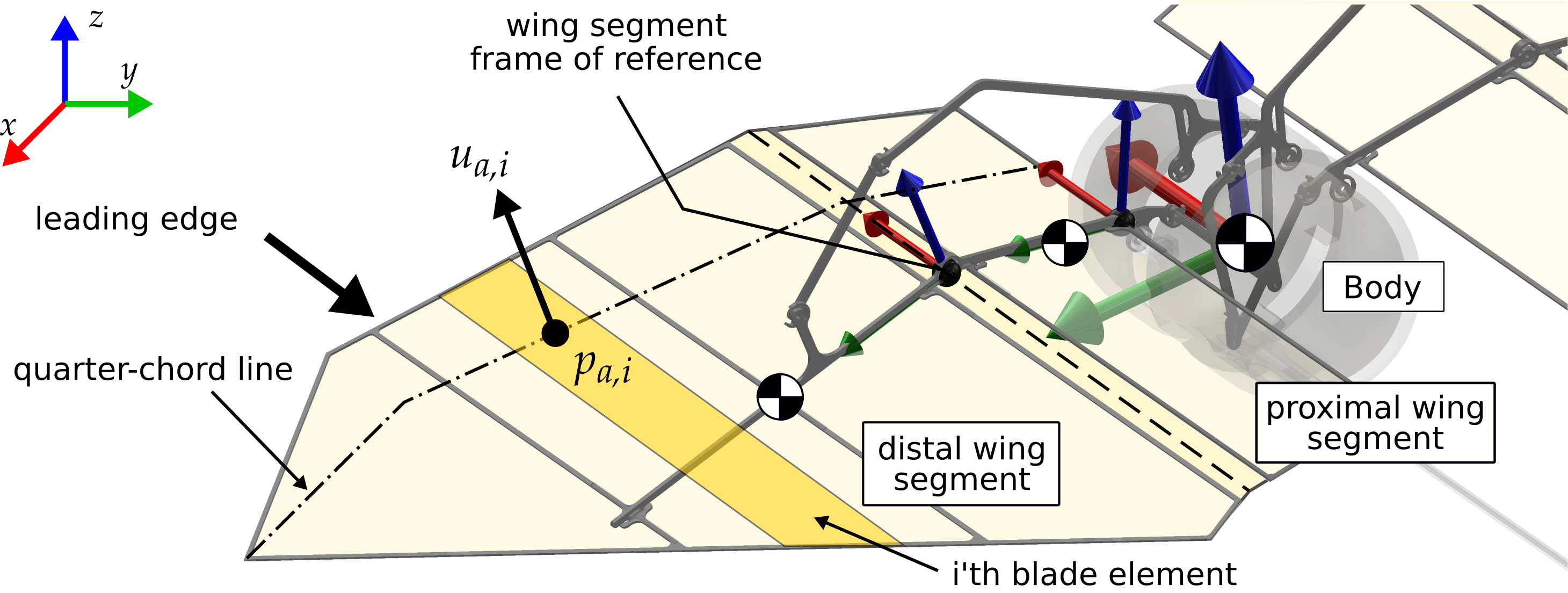}
    \caption{Illustrates quarter-chord points used to compute $B_2(q)u_2(q,\dot q)$.}
    \label{fig:aero_strips}
\end{figure*}

The aerodynamic forces are modeled by obtaining the resultant lift and drag forces $u_2$ on $n$ discrete blade elements at their quarter-chord location denoted by $p_{a,i} \in \mathbb{R}^3$ ($i \in \{1, \dots, n\}$) shown in Fig.~\ref{fig:aero_strips}. The combined aerodynamic generalized forces can be derived as follows:
\begin{equation}
  B_2(q) u_2 = \left (\frac{\partial P}{ \partial q} \right)^\top u_2,
\label{eq:aero_generalizedforces}
\end{equation}
\noindent where $P = [p_1,\dots,p_i]^\top$ embodies all of the quarter-chord points, see Fig.~\ref{fig:aero_strips}. We model $B_2(q)u_2(q,\dot q)$, which is a skinny vector containing all of the aerodynamic forces at each wing section, using models of form given in \cite{izraelevitz_state-space_2017}. 

In this work, our objective is two-pronged and include 

\begin{itemize}
    \item First, we show that using algorithmic differentiation it is possible to generated models of articulated flapping flight useful for real-time model-based flight control. 
    
    \item Second, we demonstrate that iterative model estimations for $B_2(q) \, u_2(q,\dot q)$ can be achieved in a feasible way (from a closed-loop-feedback-design standpoint), allowing the accurate prediction of nonlinear fluid-structure interactions in real-time. 
\end{itemize}

\subsection*{Algorithmic Differentiation of Inertial, Coriolis Effects, Gravity in Morphing Wings}
In Eq. ~\ref{eq:lagrangian_formulation}, the inertial matrix, $D(q)$, Coriolis ,$C(q,\dot q)$, gravity vector, $G(q)$, are computed based on a precise modeling scheme that considers the joint angles, $q$, their velocities, $\dot q$, and applies forward and velocity kinematics to obtain position and velocity terms symbolically. We obtain these terms symbolically using Matlab's Symbolic Toolbox. Following standardized steps $D$, $C$ and $G$ are obtained. However, these symbolic matrices are massive and their application for fast flight control and motion planning is infeasible. 

The alternative to this is Algorithmic Differentiation (AD) \cite{carius_trajectory_2018} which is as accurate as symbolically generated results but much leads to much faster computation overhead. There are two approaches to do AD: (i) operator overloading or (ii) source code transformation. We have decided to use the second approach using the toolbox called CasADi because of its great success in path-planning and motion control in robot manipulators and floating-base legged systems. CasADi uses a powerful parser to introduce the derivative terms.    

\section{Efficient Fluid-Structure Interaction Model Estimation Using Cubature Rules}

Here, our objective is to find estimated models for fluid-structure interactions in articulated flapping flight. The main motivation is to obtain models that have low computation overhead, can accurately predict aerodynamic contributions, and as a result can be used for real-time flight control. It is possible to show that these fluid-structure interactions, in addition to the state vector $x$, can be parameterized in terms of some internal variables in form of indicial models given by 

\begin{equation}
\Sigma_{Aero}:\left\{
\begin{aligned}
    \dot \xi &= A_\xi (t,x) \xi + B_\xi (t,x) h(x) \\\vspace{0.5cm}
    h_2 &= C_\xi (t,x)\xi + D_\xi (t,x) h(x)
\end{aligned}
\right.
\label{eq:ss-rep-aerodyn}
\end{equation}

\noindent where $A_\xi$, $B_\xi$, $C_\xi$ and $D_\xi$ are some state- and time-dependent matrices. The hidden variables are denoted by $\xi$, and $h(x)$ defines the holonomic constraints that describe morphing in the system (enforced by $u_1$ from Eq.~\ref{eq:lagrangian_formulation}). The form of Eq.~\ref{eq:ss-rep-aerodyn} has motivated us to use neural network policies to find aerodynamic models described by

\begin{equation}
\Sigma'_{Aero}:\left\{
\begin{aligned}
    \dot \xi &= \gamma_1(x) \\\vspace{0.5cm}
    a &= \gamma_2(x,\xi)
\end{aligned}
\right.
\label{eq:fsi_model}
\end{equation}

\noindent where $\gamma_1$ is the state-dependent, internal dynamics capturing fluid-structure interactions and $\gamma_2$ is a nonlinear function of the hidden variable $\xi$. Note that Eq.~\ref{eq:fsi_model} partly captures the behavior of Eq.~\ref{eq:ss-rep-aerodyn} as it takes the state vector as the input and ignores $\xi$'s role in steering $\dot \xi$. So, the idea is to capture this hidden dynamics using neural net policies. In other words, throughout this section our main objective is to sample the robot's configurations (i.e., $x=[q^\top,\dot q^\top]^\top$) at $m$ time points during flight and utilize vehicles Lagrangian dynamics, which present the predictable part of flight dynamics, to train a deep neural network $\phi$ such that an error norm is minimized, which will be briefly explained later. 

The deep neural network is a hierarchical model of multiple layers and activation functions, where each layer applies a linear transformation followed by a nonlinear activation function $\psi$ to the preceding layer. The equation of $\phi$ is given by

\begin{equation}
    \phi(X,W^1,\dots,W^i) = \psi_i(\psi_{i-1}(\psi_1(XW^1)\dots W^{i-1})W^i)
    \label{eq:neural_net}
\end{equation}

\noindent Where $W^i$ is the vector containing the weights of i-th layer and $X = [x_1,\dots,x_m]$ embodies the sampled robot configurations. The activation function at i-th layer is denoted by $\psi_k$. So, the general idea is to use sampled robot configurations to construct training data $Y = [a_1,\dots,a_m]$ and then use that to train the neural net weights $W$ such that the error norm given below is minimized 

\begin{equation}
\mathcal{W}^*= \operatorname*{argmin}_{\mathcal{W}}\left\Vert Y- \phi(q,\dot q,W^1,\dots,W^i)\right\Vert.
\label{eq:err_norm}
\end{equation} 

\noindent where $\mathcal{W}^*=[W^{1*},\dots,W^{i*}]$ is the optimal neural network weight. Note that the training data $a_j$ (j denotes sample index) is directly obtained from the model dynamics as

\begin{equation}
\begin{aligned}
B_2(q_j) \, u_2(q_j,\dot q_j)&=D(q_j) \ddot q_j + C(q_j, \dot q_j)\dot q_j+\\
&G(q_j) - B_1(q_j) \, u_1(q_j,\dot q_j)
\end{aligned}
\label{eq:training_data}
\end {equation}

\noindent where $q_j$ and $\dot q_j$ denote robot configurations at j-th sample time. Backpropagation is usually used to find the optimal weights. Since backproppagation requires large amount of time to find $\mathcal{W}^*$, many samples (i.e., skinny $Y$ matrix) have to be collected offline and the optimization process have to also be done offline. 

Another problems with backprogagation involves narrow boundaries of prediciton which becomes very important in our application. Meaning to realistically predict fluid-structure interactions, the measurements must be collected at various robot configurations which is not feasible. As a result, we can end up with measurements collected and weights trained around a local attractor. There is the risk that when the robot states escape that local attractor, neural network weights $\mathcal{W}$ cannot be updated in real-time due to the computational cost of backproppagation.

Extended Kalman Filter (EKF) can be considered to estimate $\mathcal{W}^*$ \cite{ljung_asymptotic_1979}. Since in the neural network case, EKF method depends on computing the Jacobian term $\frac{\partial \phi}{\partial \mathcal{W}}$ the computational cost can be very high. In this paper, we use cubature numerical approximation rules to find the optimal weights $\mathcal{W}^*$. Unlike the EKF \cite{ljung_asymptotic_1979}, the cubature rules do not need to solve the jacobian term $\frac{\partial \phi}{\partial \mathcal{W}}$, and can offer practical numerical solutions for computing the Gaussian-weighted integrals that appear in posterior terms involved in model estimation. 

\subsection{Model Estimation Problem}

In general, the KF training algorithm can be divided into two classes, parallel KF and parameter-based KF. In parallel KF, both neural network output and weights are the states to be estimated by the KF algorithm. For parameter-based KF, only weights are treated as the states to be estimated. We consider the second approach which treats the training process as a filtering process.

We will turn the problem of finding the unknown internal dynamics given by Eq.~\ref{eq:fsi_model} into the following neural network model estimation problem. Consider the neural network weights to be found in Eq.~\ref{eq:err_norm}. The following discritized dynamics are considered

\begin{equation}
\Sigma_{NN}:\left\{
\begin{aligned}
W^1_{k+1}&=W^1_{k}+\nu^1_k\\
W^2_{k+1}&=W^2_{k}+\nu^2_k\\
\vdots\\
W^{i-1}_{k+1}&=W^{i-1}_{k}+\nu^{i-1}_k\\
W^i_{k+1}&=W^i_{k}+\nu^i_k\\
a_{k+1}&=\phi(x_{k+1},W_{k+1}^1,\dots,W_{k+1}^i)+r_k
\end{aligned}
\right.
\label{eq:nn_fsi_model}
\end{equation}

\noindent where the subscripts denote the k-th sample time; $\nu^i_k$ and $r_k$ are the process and measurement noise, respectively. We stack all of the weight parameters and associated noise from i-th layers to simplify the notation as following

\begin{equation}
\Sigma'_{NN}:\left\{
\begin{aligned}
\mathcal{W}_{k+1} &= \mathcal{W}_{k} + \nu_k\\ 
a_{k+1}&=\phi(x_{k+1},\mathcal{W}_{k+1})+r_k
\end{aligned}
\right.
\label{eq:compact_nn_fsi_model}
\end{equation}

\noindent where $\mathcal{W}_k$ and $\nu_k$ embody the stacked parameters from each layer. Note that the neural network weights are the state vector and the vehicle configuration $x$ is the input in Eq.~\ref{eq:compact_nn_fsi_model}. While, in Eq.~\ref{eq:fsi_model}, we assumed the output $\gamma_2$ is a function of states $x$ and hidden variable $\xi$, in Eq.~\ref{eq:compact_nn_fsi_model}, we assume the output function $\phi$ is the function of states and neural net weights ($\mathcal{W}$) which explains how the hidden dynamics is replaced by neural net policies to be found in this section. 

We also note that in Eq.~\ref{eq:compact_nn_fsi_model}, the main reason that the process dynamics is simply identity is because the estimation problem is solved under the assumption that the weights are optimal weights in which case their value must remain unchanged, i.e., $\mathcal{W}_{k+1}=\mathcal{W}_k$. Ideally, $\mathcal{W}_k$ must be constant. However, we add the artificial process noise $\nu_k$ to provide more flexibility in tuning the filter. 

\subsection{Bayesian Estimation}

The general form for the estimated dynamics of Eq.~\ref{eq:compact_nn_fsi_model} is given by

\begin{equation}
\Sigma_{KF}:\left\{
\begin{aligned}
\hat{\mathcal{W}}_{k+1}&=\hat{\mathcal{W}}_k+K_k(\hat{a}_k-B_2u_2)\\
\hat{a}_{k+1}&=\phi(x_{k+1},W_{k+1}^1,\dots,W_{k+1}^i)+r_k
\end{aligned}
\right.
\label{eq:estim_nn_fsi_model}
\end{equation}

\noindent where $\hat{\mathcal{W}}_k$ is the estimated weight vector at k-th time step, $K_k$ is the generated Kalman gain according to cubature approximation rules inspired by \cite{arasaratnam_cubature_2009-1}, and $\hat{a}_k$ is the output of the neural network. 

Assume we have a set of measurements $A_k = [a_1,a_2,\dots,a_k]$, the goal of any bayesian estimator is to estimate state $\mathcal{W}_{k+1}$ form measurements $A_{k+1}$ by finding the posterior density function 

\begin{equation}
    p(\mathcal{W}_{k+1}|A_{k+1})=\frac{p(a_{k+1}|\mathcal{W}_{k+1})}{p(a_{k+1}|A_k)}p(\mathcal{W}_{k+1}|A_k)
    \label{eq:posterior_function}
\end{equation}

\noindent where $p(.)$ denotes the probability density function and its fully determined by it is mean value and covariance matrix. The posterior density function $p(W_{k+1}|A_{k+1})$ is fully determined by obtaining the mean and covariance matrix of $p(a_{k+1}|\mathcal{W}_{k+1})$, $p(a_{k+1}|A_k)$ and $p(\mathcal{W}_{k+1}|A_k)$. Since finding the mean and covariance terms for all of these terms follow a similar procedure, here, let us only show obtaining the mean value $\hat{\mathcal{W}}_{k+1|k}$ and covariance matrix $P_{k+1|k}$ of $p(\mathcal{W}_{k+1}|A_k)$. 

The expected value of $\mathcal{W}$ given the \textit{a priori} information $A_k$ is given by the following generic equation

\begin{equation}
\hat{\mathcal{W}}_{k+1|k} = \int_{\mathcal{D}} \mathcal{W}_kp(\mathcal{W}_{k}|A_k)d\mathcal{W}_k
\label{eq:mean_w}
\end{equation}

\noindent where $\mathcal{D}$ is the space on which $\mathcal{W}$ is defined. Note that to write Eq.~\ref{eq:mean_w}, Eq.~\ref{eq:compact_nn_fsi_model} and the fact the $\nu_k$ is zero-mean random variable are used. In Eq.~\ref{eq:mean_w}, $p(\mathcal{W}_{k}|A_k)=\mathcal{N}(\mathcal{W}_k;\hat{\mathcal{W}}_{k|k},P_{k|k})$ is the Gaussian density function with the mean value given by $\hat{\mathcal{W}}_{k|k}$ and covariance matrix given by $P_{k|k}$. The covariance matrix $P_{k+1|k}$ is given by

\begin{equation}
\begin{aligned}
P_{k+1|k} &=\int_{\mathcal{D}} \mathcal{W}_k\mathcal{W}^T_k \mathcal{N}(\mathcal{W}_k;\hat{\mathcal{W}}_{k|k},P_{k|k})d\mathcal{W}_k -\\ 
&\hat{\mathcal{W}}_{k+1|k}\hat{\mathcal{W}}_{k+1|k}^T +Q_{k}
\end{aligned}
\label{eq:cov_w}
\end{equation}

\noindent where $Q_k$ is the process noise covariance matrix at k-th sample time. Next, we briefly describe the approximation methods used to find Guassian-weighted integrals in the above equations. 

\subsection{Approximation of Probability Density Terms}

When the predictive (process) and observation models are linear, that is the Gaussian-weighted integrals to be solved possess the form

\begin{equation}
    I(l)=\int_{\mathcal{D}_x}l(x)p(x)dx
\end{equation} 

\noindent where $l(x)$ is a linear function of state vector $x$ and $p(x)$ is the a Gaussian density function, then, the integral given above can be solved analytically, leading to the well-known Kalman time and measurement update equations\cite{calandra_bayesian_2016,rai_bayesian_2017,ho_bayesian_1964}. 

When the predictive and observation models are nonlinear as is the case presented by Eq.~\ref{eq:nn_fsi_model}, the linearization of the model \cite{ljung_asymptotic_1979} or numerical approximation solutions must be implemented. For instance, EKF linearizes the model using Taylor series expansion. The linezarization process can be computationally expensive for the application of Eq.~\ref{eq:nn_fsi_model} in real-time closed-loop feedback flight control of our models. Mainly because, EKF heavily depends on $\frac{\partial \phi}{\partial \mathcal{W}}$ at every time step. To solve this issue, we consider filters that use cubature rules to compute the Guassian-weighted integrals that appear in the conditional probability density function explained previously. 

Consider the following nonlinear Gaussian-weighted integral, which captures terms such as Eq.~\ref{eq:cov_w}, 

\begin{equation}
  I(f)=\int_{\mathcal{D}} f(\mathcal{W})exp(-\mathcal{W}^T\mathcal{W})d\mathcal{W} 
  \label{eq:integral_exp}
\end{equation}

\noindent where $f(.)$ denotes any nonlinear function of $\mathcal{W}$. Then, the integral given by Eq.~\ref{eq:integral_exp}, can be approximated by

\begin{equation}
    I(f)=\sum_{i=1}^m b_i f(\xi_i)
    \label{eq:approx_integral_exp}
\end{equation}

\noindent where $m=\text{sample number}$, $b_i=\frac{1}{m}$ and $\xi_i=\sqrt{\frac{m}{2}}[1]_i$. Let $w_1(\mathcal{W})=\mathcal{N}(\mathcal{W};\mu,\Sigma)$ (where $\mu$ and $\Sigma$ are ...) and $w_2(\mathcal{W})=exp(-\mathcal{W}^T\mathcal{W})$, then it is possible to show that

\begin{equation} 
\int_{\mathcal{D}} f(\mathcal{W})w_1(\mathcal{W})d\mathcal{W}=\frac{1}{\sqrt{\pi^n}}\int_{\mathcal{D}} f(\sqrt{2\Sigma}\mathcal{W} +\mu)w_2(\mathcal{W})d\mathcal{W}
\label{eq:cubature_rule}
\end{equation}

\noindent From Eqs.~\ref{eq:integral_exp}, \ref{eq:approx_integral_exp} and \ref{eq:cubature_rule} it can be shown that

\begin{equation}
    \int_{\mathcal{D}} f(\mathcal{W})\mathcal{N}(\mathcal{W};\mu,\Sigma)d\mathcal{W}=\sum_{i=1}^m w_if(\sqrt{2\Sigma}\xi_i +\mu)
\end{equation}

\noindent which is directly used to write the algorithm given below to compute the approximation of Gaussian-weighted integrals in the Kalman Filter.

\begin{algorithm}
\begin{algorithmic}[h]
\caption{Compute Cubature Kalman Gain and Covariance Matrix} 

\Function{Update Kalman Gain}{}

\State $P_{k-1|k-1} \gets S_{k-1|k-1}S_{k-1|k-1}^{T}$ 

\vspace{0.25cm}

\textbf{Predictive Update:}

\State $\mathcal{W}_{i,k-1|k-1}\gets S_{k-1|k-1}\xi_{i} + \hat{\mathcal{W}}_{k-1|k-1}$ 

\State $\mathcal{W}_{i,k|k-1}^{*} \gets \mathcal{W}_{i,k-1|k-1}$

\State $\hat{\mathcal{W}}_{k|k-1} \gets \frac{1}{m}\sum_{i=1}^{m}\mathcal{W}_{i,k|k-1}^{*}$

\vspace{0.25cm}

\State $P_{k|k-1}\gets\frac{1}{m}\sum_{i=1}^{m}\mathcal{W}_{i,k|k-1}^{*}\mathcal{W}_{i,k|k-1}^{*T}$
\State $\hspace{1.0cm}-\hat{\mathcal{W}}_{k|k-1}\hat{\mathcal{W}}_{k|k-1}^{T} +Q$

\vspace{0.25cm}

\textbf{Measurement Update:}

\State $\mathcal{W}_{i,k|k-1}\gets S_{k|k-1}\xi_{i} + \hat{\mathcal{W}}_{k|k-1} $

\State $Y_{i,k|k-1}\gets \phi(X_{i,k},\mathcal{W}_{i,k|k-1})$

\State $\hat{a}_{k|k-1}\gets\frac{1}{m}\sum_{i=1}^{m}Y_{i,k|k-1}$

\vspace{0.5cm}

\State $P'_{k|k-1}\gets\frac{1}{m}\sum_{i=1}^{m}Y_{i,k|k-1}Y_{i,k|k-1}^{T}$

\State $\hspace{1cm} -\hat{a}_{k|k-1}\hat{a}_{k|k-1}^{T} +R$

\vspace{0.5cm}

\State $P''_{k|k-1}\gets\frac{1}{m}\sum_{i=1}^{m}\mathcal{W}_{i,k|k-1}Y_{i,k|k-1}^{T}-\hat{\mathcal{W}}_{k|k-1}\hat{a}_{k|k-1}^{T} $

\State $ K_{k}\gets P''_{k|k-1}\left(P'_{k|k-1}\right)^{-1} $

\State $ P_{k|k}\gets P_{k|k-1}-K_kP'_{k|k-1}K_k^{T}$

\vspace{0.25cm}

\textbf{return} $P_{k|k},~K_k$

\EndFunction

\end{algorithmic}
\end{algorithm}

\section{Simulation Results and Discussion}

Here, we briefly report the accuracy and computation overhead of the obtained model for our morphing wing flight based on AD and cubature rules. Our results suggest the feasibility of using these models for locomotion control of robots whose dynamics involves sophisticated robot-environment interactions.

\begin{figure}[t!]
\includegraphics[width = 0.9 \linewidth]{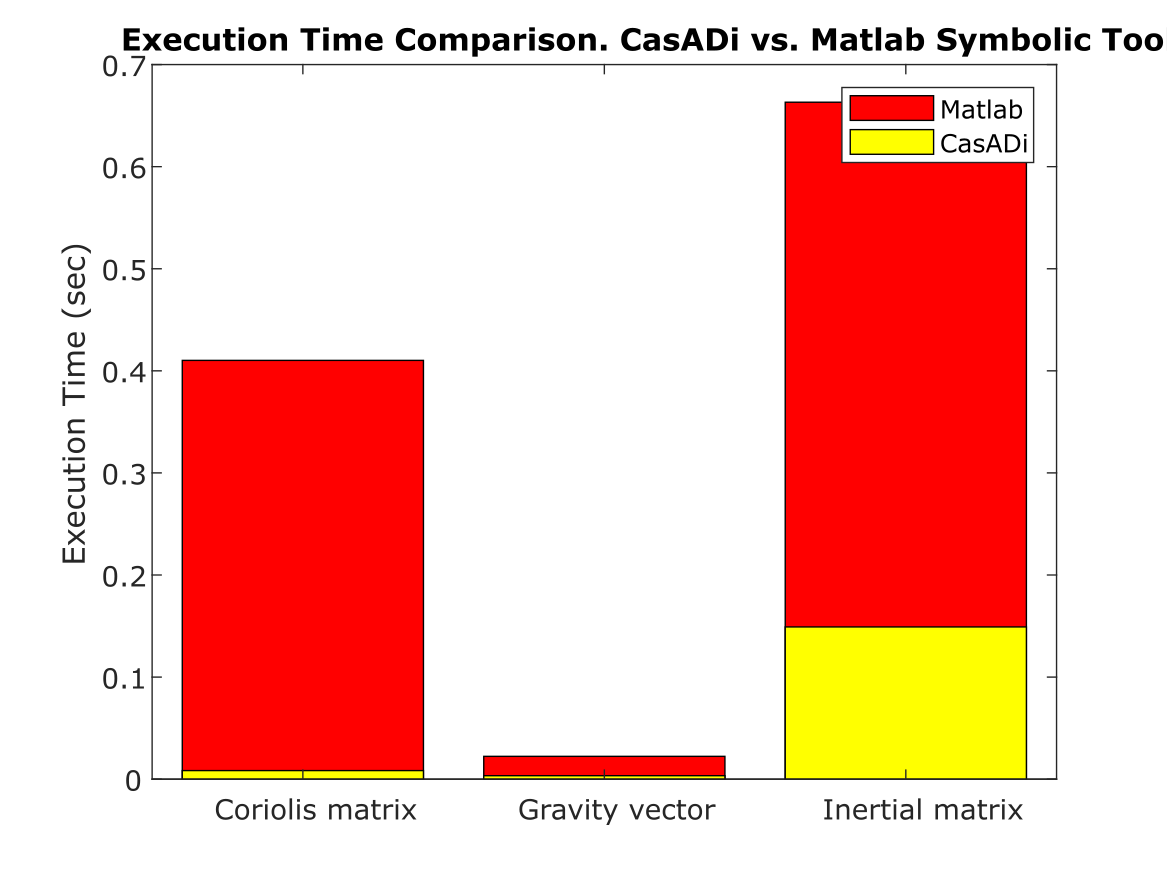} 
\caption{Shows the comparison between the computation overhead of MATLAB Symbolic Toolbox- and CasADi-generated inertial, Coriolis and Gravity matrices.} 
\label{fig:overhead-comp-dyn}
\end{figure}

\begin{figure}[t!]
\includegraphics[width = 0.9 \linewidth]{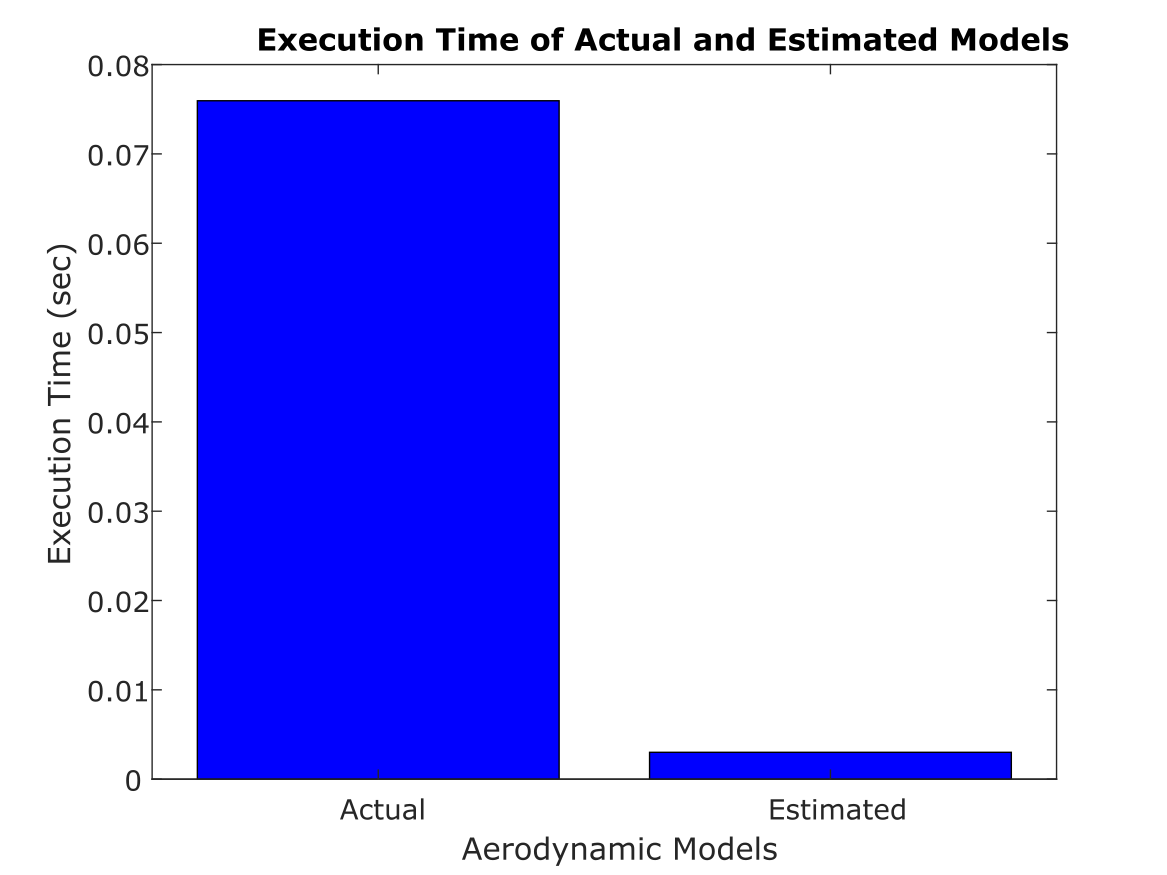} 
\caption{Shows the comparison between the computation overhead of actual and estimated aerodynamic models.} 
\label{fig:overhead-comp-Aero}
\end{figure}

Figures~\ref{fig:overhead-comp-dyn} and \ref{fig:overhead-comp-Aero} show the computation overhead for various terms in the inertial dynamics. As expected AD yields significantly smaller computation overhead. To check the accuracy of AD models versue symbolically generated terms, we use AD-based inertia, Coriollis and gravity terms to obtain training data points for our cubature filter. 

Note that we use a mathematical model to extract our training data $B_2(q)u_2(q,\dot q)$. However, the proposed framework is developed with the aim of implementing it on to the actual robot. So, without loss of generality the same training data can be obtained by capturing the vehicle's body position, orientation, wing angles and their times derivatives and using the vehicles Lagrangian dynamics (excluding aerodynamics interactions). 

\begin{figure}
\includegraphics[width = 1.0 \linewidth]{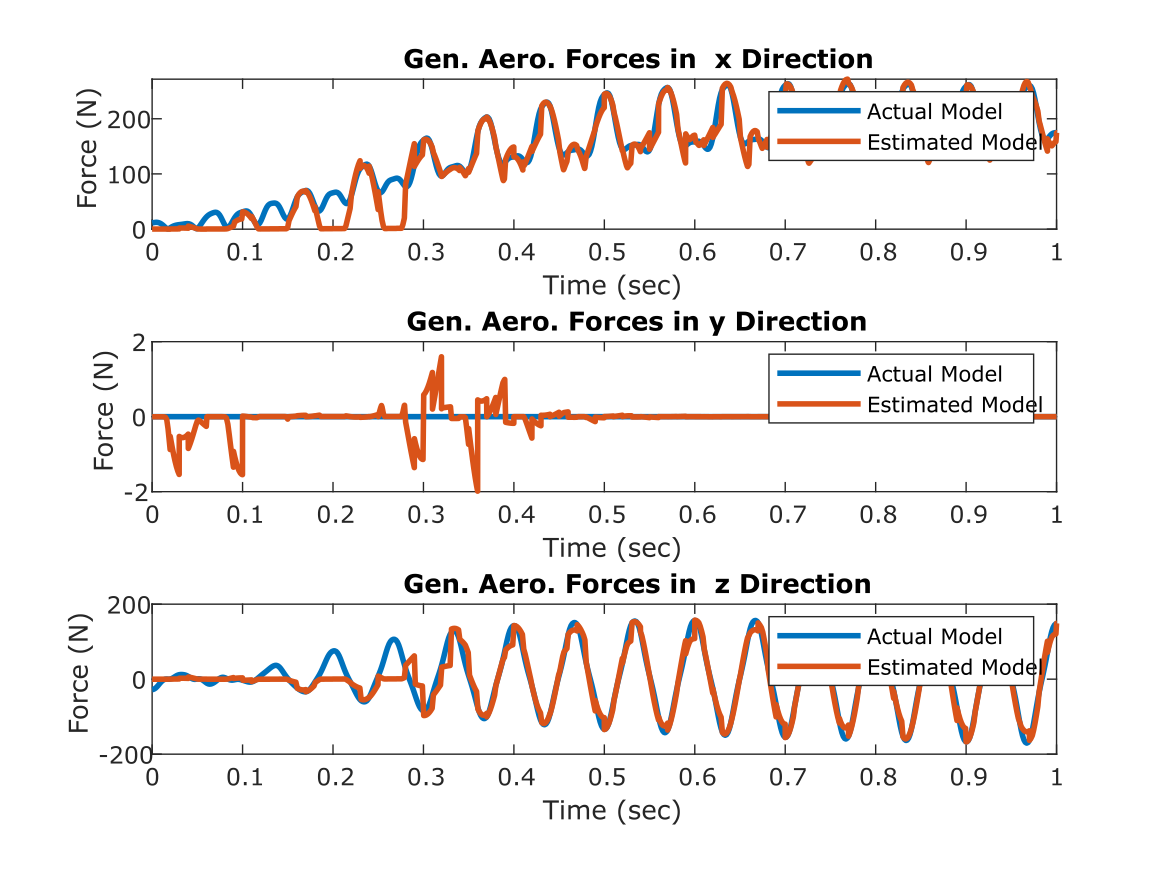} 
\caption{Shows a comparison between the outputs from the generalized forces ($B_2(q)u_2(q,\dot q)$) in the x,y and z directions from the actual and estimated aerodynamic models.} 
\label{fig:xyz_gen_force}
\end{figure}

\begin{figure}
\includegraphics[width = 1.0 \linewidth]{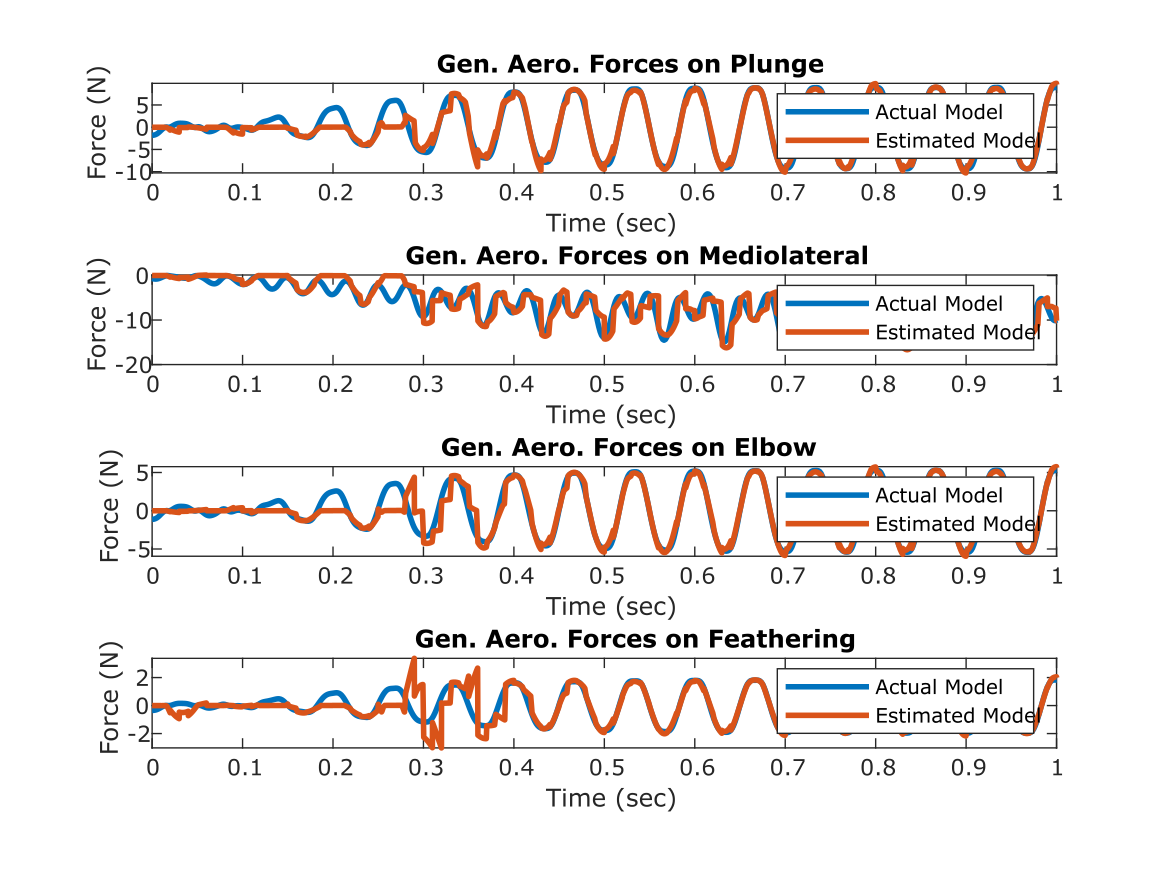} 
\caption{Shows a comparison between the outputs from the generalized forces on the left wing joints from the actual and estimated aerodynamic models (Note, left and right wing have symmetric morphing).} 
\label{fig:Left_Wing_gen_force}
\end{figure}

Figures~\ref{fig:xyz_gen_force} and \ref{fig:Left_Wing_gen_force} show a comparison between the estimated and actual models. In these simulation results, our neural network consists of three layers with two softplus activation functions for form $\psi=ln(1+\exp^{in})$. At every 100 samples, one training data is utilized towards obtaining optimal neural network weights using the cubature Kalman Filter. And then the learned models is used to estimated fluid structure interactions in the robot. 

The robots' body joint angles are commanded with predefined values. Damping coefficients are considered to passively stabilize roll, pitch, and yaw dynamics as the vehicle is open-loop unstable and no controller is involved. 

Based on our simulation results, and as it can be seen in Figs.~\ref{fig:xyz_gen_force} and \ref{fig:Left_Wing_gen_force} the estimated aerodynamics model can do a decent job predicting actual generalized forces $B_2(q)u_2(q,\dot q)$ in a 200-milliseconds time envelope (sample time is $10^{-4}$ sec). The process and measurement noise values ($r_k$ and $\nu_k$) are assumed to be zero. The mean square error is $0.04\%$ for all generalized forces. The diagonal terms in the covariance matrix $P$ are less than $10^{-2}$ which reflects the confidence of the filter about the estimated weights. 

\section{Concluding Remarks}

In this work, algorithmic differentiation and neural networks equipped with cubature rules were proposed to identify inertial and fluidic forces acting on a morphing MAV called Aerobat. These forces can be very complex to the extent that averaged models fail to predict them. The neural network design extends to prior works on Bayesian filters. Using cubature rules to compute Gaussian-weighted integrals efficiently, we showed that the complex multi-degrees-of-freedom dynamics of morphing MAVs can be computed very efficiently and accurately. In the future, we aim to integrate online fluidic structures identification of these morphing MAV with real-time closed-loop feedback control and test them experimentally on our Aerobat platform.


\bibliographystyle{bibstyle_favorite}
\bibliography{references}

\end{document}